\documentclass[aps,prl,twocolumn]{revtex4}
\usepackage{graphics}
\usepackage{graphicx}

\begin{document}


\title{Screening and localization in the nonlinear Anderson problem}




\author{Alexander~V.~Milovanov${}^{1,3}$ and Alexander~Iomin${}^{2,3}$}

\affiliation{${}^1$ENEA National Laboratory, Centro~Ricerche~Frascati, I-00044 Frascati, Rome, Italy}
\affiliation{${}^2$Solid State Institute, Technion$-$Israel Institute of Technology, 32000 Haifa, Israel}
\affiliation{${}^3$Max Planck Institute for the Physics of Complex Systems, D-01187 Dresden, Germany}




\begin{abstract} We study the spreading dynamics of an initially localized wave packet in 1D nonlinear Schr\"{o}dinger lattices with random potential. It is shown that adding small dielectric coupling to surrounding random medium results in asymptotic localization of the nonlinear field. The nonlinear localization length depends on dielectric loss of the medium at low temperatures and the value of nonlinearity parameter. The model predicts a possibility of self-induced localization when the ``medium" to which the wave field is dielectrically coupled is the nonlinear wave itself. 
\end{abstract}


\maketitle

Anderson localization \cite{And} is a peculiar form of dynamic localization, which characterizes the transport of waves in disordered systems. The physical picture is that of a falling wave scattered by impurities of the medium. For strong disorder, this scattering process occurs in multiple scattering paths along which the linear field interferes with itself, leading to a spatially localized wave function akin to a standing wave. The phenomenon was initially discussed \cite{And,Thou,PW16} for the dynamics of noninteracting electrons in a semiconductor and at present is fairly well understood \cite{50}.  

Can nonlinearity or interactions destroy Anderson localization in random lattices? The question\textemdash to which no unambiguous answer has been provided\textemdash challenges the classic theories of localization \cite{Rep}, which assume that the lattice is static and the absence of interactions. Over the years, two nonequivalent approaches to the problem and respectively two different lines of inquiry have emerged. 

One\textemdash dating back to Fleishman and Anderson \cite{Fleish}\textemdash pursued the interaction path, with a major milestone settled in by Altshuler {\it et al.} \cite{Alt}. This line of inquiry has evolved over time into a vast area of research commonly referred to as many-body localization, or MBL (Refs. \cite{MBL,Tod17,Tod18,Tod19,Roy} for reviews; references therein). The result of interest here is that electron-electron interactions alone are unable to cause the relaxation and establish the thermal equilibrium, that is one needs a thermal bath (coupling to phonons) to destroy the localized state \cite{MBL}. 

The other line of inquiry\textemdash initiated by Pikovsky and Shepelyansky \cite{PS} and Flach {\it et al.} \cite{Flach}, and which is the main interest of the present study\textemdash followed the nonlinearity path \cite{Sh87,Sh93}, with most attention paid to the spreading dynamics of an initially localized wave packet of finite norm in 1D nonlinear Schr\"{o}dinger lattices with random potential. The process is described by the discrete Anderson nonlinear Schr\"odinger equation (DANSE)
\begin{equation}
i\hbar\frac{\partial\psi_n}{\partial t} = \hat{H}_L\psi_n + \beta |\psi_{n}|^{2}\psi_n,
\label{DANSE} 
\end{equation}
where $\psi_n = \psi (n, t)$ is complex wave function which depends on time $t$ and the position coordinate $n$,
\begin{equation}
\hat{H}_L\psi_n = \varepsilon_n\psi_n + V (\psi_{n+1} -2\psi_n + \psi_{n-1})
\label{2} 
\end{equation}
is the Anderson Hamiltonian in tight-binding approximation \cite{And}, $V$ is hopping matrix element, $\beta > 0$ characterizes nonlinearity, on-site energies $\varepsilon_n$ are randomly distributed with zero mean across a finite energy range, and the probability distribution is normalized to $\sum_n |\psi_n|^2 = 1$.   

The various aspects of DANSE~(\ref{DANSE}) are discussed in Refs. \cite{Skokos,Wang,Fishman,Iomin,Many,Basko,EPL,PRE14,Ivan,PRE17,Fistul,PRE23}. The results can be summarized by a critical value $\beta = \beta_c$, above which the nonlinear field can spread indefinitely along the lattice, and below which it is Anderson localized similar to the linear field. 

The correspondence between DANSE~(\ref{DANSE}) and MBL is not at all trivial. Indeed, DANSE~(\ref{DANSE}) is a single-particle problem, which mimics the behavior of an expanding Bose-Einstein condensate in 1D disorder potentials (e.g., Refs. \cite{PS,Fishman,Shapiro07,Shapiro12}). 

It is the purpose of the present Letter to demonstrate that DANSE~(\ref{DANSE}) leads to asymptotic localization of the nonlinear field for {\it any} $\beta > 0$ as soon as a small dielectric coupling to ambient medium is added. This coupling of the system and the environment introduces a nonlinear localization scale, which depends exclusively on dielectric loss of the material at low temperatures and the $\beta$ value. 

Our idea is as follows: By its formulation, DANSE~(\ref{DANSE}) refers to a wave process in a medium\textemdash not in vacuum. The effect of the medium is included via the random potential in Eq.~(\ref{2}), yet there could be a nonzero dielectric response from the environment, which, however, is omitted in~(\ref{DANSE}). We argue and confirm through the direct calculation that this response is such as to diminish (screen) the actual strength of nonlinear interaction compared to a vacuum value. The result is that the nonlinear field with screened interactions is asymptotically localized {\it regardless} of the $\beta$ value, provided the lattice is infinite and the medium is in some sense random.  

More explicitly, we start with DANSE~(\ref{DANSE}), where we interpret $\beta$ as the intrinsic value of the nonlinear field. We refer to this value as the bare value. Next we add  dielectric coupling to environment and wonder about the strength of nonlinear interaction under this coupling. We show that in the presence of randomness the nonlinearity becomes dispersive, that is the actual $\beta$ value starts to depend on the position coordinate $n$. It is the dispersive character of dielectric response from the surrounding random medium that localizes the nonlinear field in 1D independently of intrinsic $\beta$. 

The theory approach designed above is inspired \cite{Falkovich} by quantum electrodynamics, where one distinguishes between the bare (predicted by fundamental theory) electric charge, and the actually measured charge \cite{AhBe,Cao}. The latter is somewhat smaller than the former as a result of screening by short-living electron-positron pairs surrounding the electric charge in vacuum. Our analysis suggests the situation behind DANSE~(\ref{DANSE}) is just similar.  

In fact, assume for concreteness that the wave function $\psi_n$ refers to an electromagnetic wave. It is understood that the electric field of the wave perturbs the surrounding dielectric medium causing its polarization. The action of the polarization field is that it partially screens the bare electric field of the wave, with that consequence that the actual strength of nonlinear interaction is decreased by a factor of 
\begin{equation}
{\Delta\beta}/{\beta} \simeq - {P_\omega}/{D_\omega},  
\label{Delta} 
\end{equation} 
where $\beta$ is the bare value, and $\Delta\beta$ is a variation due to polarization. In the above, $P_\omega$ is the amplitude of the polarization field, and $D_\omega$ is the electric displacement, both frequency dependent in general. In a basic theory of dielectrics one writes the polarization and displacement fields respectively as $P_\omega = \varepsilon_0 \chi E_\omega$ and $D_\omega \equiv \varepsilon_0 E_\omega + P_\omega$, where $\varepsilon_0$ is the permittivity of free space, and $\chi$ is the dielectric susceptibility. Upon substituted in Eq.~(\ref{Delta}) this yields $\Delta \beta / \beta \simeq -\chi/ (1 + \chi)$, from which the effective control parameter is $\beta_{\rm eff} = \beta + \Delta \beta \simeq \beta / (1+\chi)$ \cite{Charge}. 

It is understood that $\chi$ is a complex quantity, i.e., $\chi = \chi_1 + i \chi_2$, whose components can depend on the frequency of the applied field. These frequency-dependent components are related to each other via the Kramers-Kronig (Hilbert) transform, i.e., $\chi_1 (\omega) = (1/\pi) {\rm P} \int \chi_2 (\omega^\prime)d\omega^\prime/(\omega^\prime-\omega)$. In this context, one writes the effective control parameter as $\beta_{\rm eff} = \beta / [1+\chi_1 (\omega)]$, where $\chi_1 (\omega) \equiv {\rm Re}\, \chi (\omega)$ is the real part of $\chi (\omega)$. The imaginary part $\chi_2 (\omega) \equiv {\rm Im}\, \chi (\omega)$ is expressed as $\chi_2 (\omega) = \sigma_{\rm a.c.} (\omega) / \varepsilon_0\omega$, where $\sigma_{\rm a.c.} (\omega)$ is the real part of the complex frequency-dependent a.c. (alternating-current) conductivity of the medium. We should stress that the frequency dependence of both $\chi (\omega)$ and $\sigma_{\rm a.c.} (\omega)$ is particularly relevant for random systems where spatial inhomogeneity dictates a distribution of time delays to the disordered medium’s polarization response. Putting all the various pieces together, we have  
\begin{equation}
{\rm Re}\, \chi (\omega) = \frac{1}{\pi\varepsilon_0} {\rm P} \int \frac{\sigma_{\rm a.c.} (\omega^\prime)}{\omega^\prime} \frac{d\omega^\prime}{\omega^{\prime} - \omega}. \label{KrKr} 
\end{equation}   

At this point, one needs a model of a.c. conduction for the surrounding random medium. It is understood that the charge carriers\textemdash whose displacement is held responsible for the frequency-dependent $P_\omega$\textemdash are subject to the same random potential~(\ref{2}) which scatters the wave function $\psi_n$. At low temperatures, the available charge conduction mechanisms fall into two main categories \cite{Elliott2,Elliott3}, i.e., transport of a carrier by quantum-mechanical tunneling through \cite{Poll71}, or classical hopping over \cite{Elliott1}, the potential barrier separating two localized states. These both lead to linear frequency dependence of a.c. conductivity versus frequency\textemdash the former ensures that a.c. conductivity stays finite as temperature vanishes, the latter accounts for some increase of a.c. conductivity as temperature rises \cite{Comm}. We ought to note that temperature and frequency dependence of a.c. conductivity in disordered semiconductors has been the subject of very intense theoretical and experimental treatments (e.g., Refs. \cite{Elliott2,Elliott3,Lax,Jons,Dyre}). What is important for our purposes is that the linear scaling $\sigma_{\rm a.c.} (\omega) \propto \omega$ applies universally for $T\rightarrow 0$ \cite{Dyre,Dyre93,PRB01,6KT}, where $T$ is temperature, and that it is confirmed experimentally that in the absence of measurable d.c. conductivity the a.c. counterpart behaves linearly with $\omega$ \cite{Elliott3,Dyre}.

Summarizing the above reasoning, we write the a.c. conductivity as $\sigma_{\rm a.c.} (\omega) = A \omega$, where $A$ is a coefficient, which can depend on temperature $T$, but not on the frequency $\omega$. In material science \cite{Elliott3,Jons,6KT}, it is customary to express this coefficient as $A = \varepsilon_0\varepsilon_r \tan\delta$, where $\varepsilon_r$ is relative permeability, and $\tan\delta$ is dielectric loss of the material. With these definitions, Eq.~(\ref{KrKr}) reduces to ${\rm Re}\, \chi (\omega) = \mu {\rm P} \int d\omega^\prime / (\omega^\prime - \omega)$, where we have introduced $\mu \equiv A/\pi\varepsilon_0 = (\varepsilon_r / \pi) \tan\delta$.  

In the absence of meaningful d.c. (direct-current) conductivity, the integration in Eq.~(\ref{KrKr}) can be performed from $\sim \omega$ (which is assumed to be comparable to, yet somewhat greater than the dielectric loss peak frequency, $\omega_m$) and up to the actual frequency spread of the wave packet, which we shall denote by $\Delta \omega$. Assuming a broad wave packet, with $\Delta \omega \gg \omega \gtrsim \omega_m$, from Eq.~(\ref{KrKr}) one obtains, with logarithmic accuracy, ${\rm Re}\, \chi (\omega) \simeq \mu \log |\Delta \omega / \omega|$. Translating frequencies into wave numbers using $\omega = v / n$, where $v$ is the speed of the wave, one gets ${\rm Re}\, \chi (n) \simeq \mu \log |\Delta n / n|$. If, however, the spread $\Delta \omega$ is large compared to $\omega$, then the ratio $|\Delta n| / n$ is also large, i.e., $|\Delta n| \gg n$, permitting one to write the real part of $\chi$ as ${\rm Re}\, \chi (\Delta n) \simeq \mu \log |\Delta n|$, without keeping an eye on $n$. To this end, the effective $\beta$ value becomes $\beta_{\rm eff} = \beta / [1 + {\rm Re}\, \chi (\Delta n)] \simeq \beta / (1 + \mu \log |\Delta n|)$. Writing, under the logarithm sign, the spread $\Delta n$ as $|n - n^\prime|$, where $|n - n^\prime| \gg 1$, one arrives at ${\rm Re}\, \chi (n, n^\prime) \simeq \mu \log |n -n^\prime|$, from which 
\begin{equation}
\beta_{{\rm eff}} = \beta_{{\rm eff}} (n, n^\prime) \simeq \beta / (1+\mu \log |n - n^\prime|).
\label{Effe} 
\end{equation}
One sees, from Eq.~(\ref{Effe}), that the effective $\beta$ value is dispersive (depends on position coordinate $n^\prime$). To this end, the DANSE model in Eq.~(\ref{DANSE})\textemdash which assigns one single value of $\beta$ to all sites\textemdash becomes insufficient. That is one needs to employ a more general dynamical equation that would directly take into account the dispersive character of interactions taking place. Here, we work with the nonlinear Schr\"{o}dinger model with dispersive nonlinearity \cite{PRE23,EPL23}
\begin{equation}
i\hbar\frac{\partial\psi_n}{\partial t} = \hat{H}_L\psi_n + \sum_{n^\prime}\beta_{{n, n^\prime}} \frac{|\psi_{n^\prime}|^{2}}{|n - n^\prime|}\psi_n,
\label{DSPL} 
\end{equation} 
where $\beta_{{n, n^\prime}} \equiv \beta (|n-n^\prime|) > 0$ is the position-dependent nonlinearity parameter, the nonlocal kernel $\sim 1/|n-n^\prime|$ represents the Poisson potential (here associated with electrostatic interaction), and $\hat{H}_L$ is the Anderson Hamiltonian of the type given by Eq.~(\ref{2}). Note that the nonlocal term in Eq.~(\ref{DSPL}) is exact in one-particle limit and does not assume any further approximation such as, for instance, the dipolar approximation \cite{DDI} or any other approximation alike \cite{DDI2}. Note, also, that Eq.~(\ref{DSPL}) differs from the familiar Schr\"odinger-Newton equation (SNE) \cite{Penrose1,Penrose2,Penrose3,Moroz,Bahrami} in that the SNE invokes neither randomness nor dispersive interactions. Identifying, in dynamical Eq.~(\ref{DSPL}), the position-dependent $\beta_{{n, n^\prime}}$ with the effective value in Eq.~(\ref{Effe}), one obtains   
\begin{equation}
i\hbar\frac{\partial\psi_n}{\partial t} = \hat{H}_L\psi_n + \sum_{n^\prime}\frac{\beta}{1+\mu \log |n - n^\prime|}\frac{|\psi_{n^\prime}|^{2}}{|n - n^\prime|}\psi_n.
\label{DSPLOG} 
\end{equation} 
If $\mu \ll 1$, i.e., the medium is a poor conductor, then one can substitute the logarithmic form $1+\mu \log |n - n^\prime|$ with the algebraic power $|n - n^\prime|^\mu$, i.e., $1+\mu \log |n - n^\prime| \approx |n - n^\prime|^\mu$, leading to   
\begin{equation}
i\hbar\frac{\partial\psi_n}{\partial t} = \hat{H}_L\psi_n + \beta \sum_{n^\prime}\frac{|\psi_{n^\prime}|^{2}}{|n - n^\prime|^{1+\mu}}\psi_n,
\label{DSPAL} 
\end{equation} 
which is equivalent to some modification of the Poisson potential in the medium. For a broad wave packet, characterized by $\Delta \omega \gg \omega$, the model in Eq.~(\ref{DSPAL}) simplifies to a superquadratic modification of DANSE~(\ref{DANSE}), i.e.,   
\begin{equation}
i\hbar\frac{\partial\psi_n}{\partial t} = \hat{H}_L\psi_n + \beta |\psi_{n}|^{2s}\psi_n,
\label{DSPSQ} 
\end{equation} 
where $s = 1+\mu$. In fact, if the nonlinear field is spread over a large number of states $\Delta n \gg 1$, and if the variation is so smooth that $|\psi_{n+1} - \psi_{n-1}| / 2|\psi_{n}| \ll 1$, then one can replace the actual separation $|n - n^\prime|$ in Eq.~(\ref{DSPAL}) with the characteristic spread $\Delta n$, take $1/|\Delta n|^{1+\mu}$ out of the summation sign, and sum over $n^\prime$ using the conservation of the total probability $\sum_{n^\prime} |\psi_{n^\prime}|^{2} = 1$. On the other hand, if $\Delta n \gg 1$, then $\sum_{n^\prime} |\psi_{n^\prime}|^{2} = 1$ dictates $|\psi_{n}|^{2} \sim 1/\Delta n$, from which $1/|\Delta n|^{1+\mu} \sim |\psi_n|^{2(1+\mu)}$. Eliminating $\Delta n$ from the nonlinear term with the aid of $\beta/|\Delta n|^{1+\mu} \sim \beta |\psi_n|^{2(1+\mu)}$, one is led to Eq.~(\ref{DSPSQ}), where $s = 1+\mu > 1$.

Equation~(\ref{DSPSQ}) predicts that the nonlinear field can spread along the lattice up to a certain finite distance only, beyond which it is Anderson localized similar to the linear field. This is because the nonlinear term in Eq.~(\ref{DSPSQ}) acts as a stochastic pump \cite{PS,HH} whose strength decays while spreading \cite{PRE14,DNC}. In fact, if the nonlinear field is spread over $\Delta n$ states, then the nonlinear frequency shift is $\Delta\omega_{\rm NL} \simeq \beta |\psi_n|^{2s} \simeq \beta/|\Delta n|^s$, while the distance between the eigenstates approaches zero as $\delta\omega \sim 1/|\Delta n|$. Hence it follows that the Chirikov overlap parameter \cite{CZ,Sagdeev} decays with $\Delta n$ as $K \equiv \Delta\omega_{\rm NL} / \delta\omega \simeq \beta/|\Delta n|^\mu$. If $K \gg 1$, then the nonlinear field in Eq.~(\ref{DSPSQ}) behaves as a chaotic system, meaning the integrals of motion are destroyed, and the nonlinear field can expand in the $n$ direction via a stochastic process. On the contrary, if $K \ll 1$, then the behavior is regular, and the field is localized. Assume the nonlinear field is initially so narrow that $K \gg 1$. Then it starts to spread along the lattice driven by the nonlinear term. On the other hand, because $K$ decreases with $\Delta n$ increasing, this spreading process comes to a halt as soon as $K \sim 1$. Further spreading is forbidden by the transition to regularity and the $K$ values becoming smaller than 1. One sees that the nonlinear field automatically (without tuning of parameters) evolves into a marginally regular state at the border of chaos and saturates exactly at that border. By the time regularity is reached the field has spread over $|\Delta n|_{\max} \sim \exp [(1/\mu) \ln \beta]$ states. In a 1D model, this is equivalent to a nonlinear localization scale  
\begin{equation}
\Lambda_{\rm loc} \simeq \exp [(\pi / \varepsilon_r \tan\delta) \ln \beta],
\label{LocL} 
\end{equation} 
where $\tan\delta$ is dielectric loss tangent. If losses are absent, then $\tan\delta$ vanishes, leading to $\Lambda_{\rm loc}\rightarrow+\infty$. That means that the nonlinear field without dielectric coupling to environment can spread to arbitrarily long spatial scales, provided $K \gg 1$. This recovers the type of dynamics considered previously \cite{PS,Flach,Skokos,Many,EPL} in the framework of DANSE~(\ref{DANSE}). Remark that the dependence of $\Lambda_{\rm loc}$ on $\mu\rightarrow +0$ is nonperturbative, as it should \cite{EPL,PRE14}.

Let us now explore a possibility that the ``medium" to which the wave field is dielectrically coupled is the nonlinear field itself. The problem\textemdash which makes sense for broad wave packets\textemdash leads to a phenomenon that one may characterize as self-screening. The self-screening occurs because the nonlinearity $\sim\beta|\psi_n|^2$ causes the components of the wave field to overlap. This destabilizes the wave function at some locations, around which the wave field turns into a chaotic state. If such states appear sufficiently close to each other, then the nonlinear field can propagate between the different chaotic states by either tunneling through, or hopping over, the regular states in-between. The process is similar to the transport mechanism considered by Elliott \cite{Elliott2,Elliott3}, with chaotic states thought of as the analog localized states. 

More explicitly, in the basis of linearly localized modes, the evolution of the components $a_k$ of the wave field is due to their nonlinear coupling, that is $i\dot{a}_k \sim \beta a_{m_1} a^*_{m_2} a_{m_3}$, where star denotes complex conjugate, the upper dot means time differentiation, $a_{-m} = a^*_m$ by way of temporal translation symmetry, and we have set $\hbar = 1$ for simplicity. If all couplings are next-neighbor-like (a situation of prime interest here as such couplings have the largest cross-section), then the evolution equation for $a_k$ reduces to (for symmetry reasons) 
\begin{equation}
i\dot{a}_k \sim 2\beta a_{k -1} a^*_{k} a_{k +1}.
\label{9} 
\end{equation} 
If $k\rightarrow\pm\infty$, then the set of dynamical equations~(\ref{9}) defines an infinite chain process with the topology of a Cayley tree. This is shown by constructing a one-to-one correspondence between~(\ref{9}) and a Cayley graph as follows. Each cyclic term $i\dot{a}_k$ with the wave number $k$ and the eigenfrequency $\omega_k = \omega (|k|)$ is represented as a node in the graph space labeled by the coordinate $k$. The coupling links between the different nodes are represented as bonds. From the structure of the nonlinear term in Eq.~(\ref{9}) one sees there will be three and only three bonds at each node\textemdash one ingoing bond representing the wave process with amplitude $a^*_{k}$, and two outgoing bonds representing the processes with amplitudes $a_{k-1}$ and $a_{k+1}$, respectively (see Fig.~1).

\begin{figure}
\includegraphics[width=0.52\textwidth]{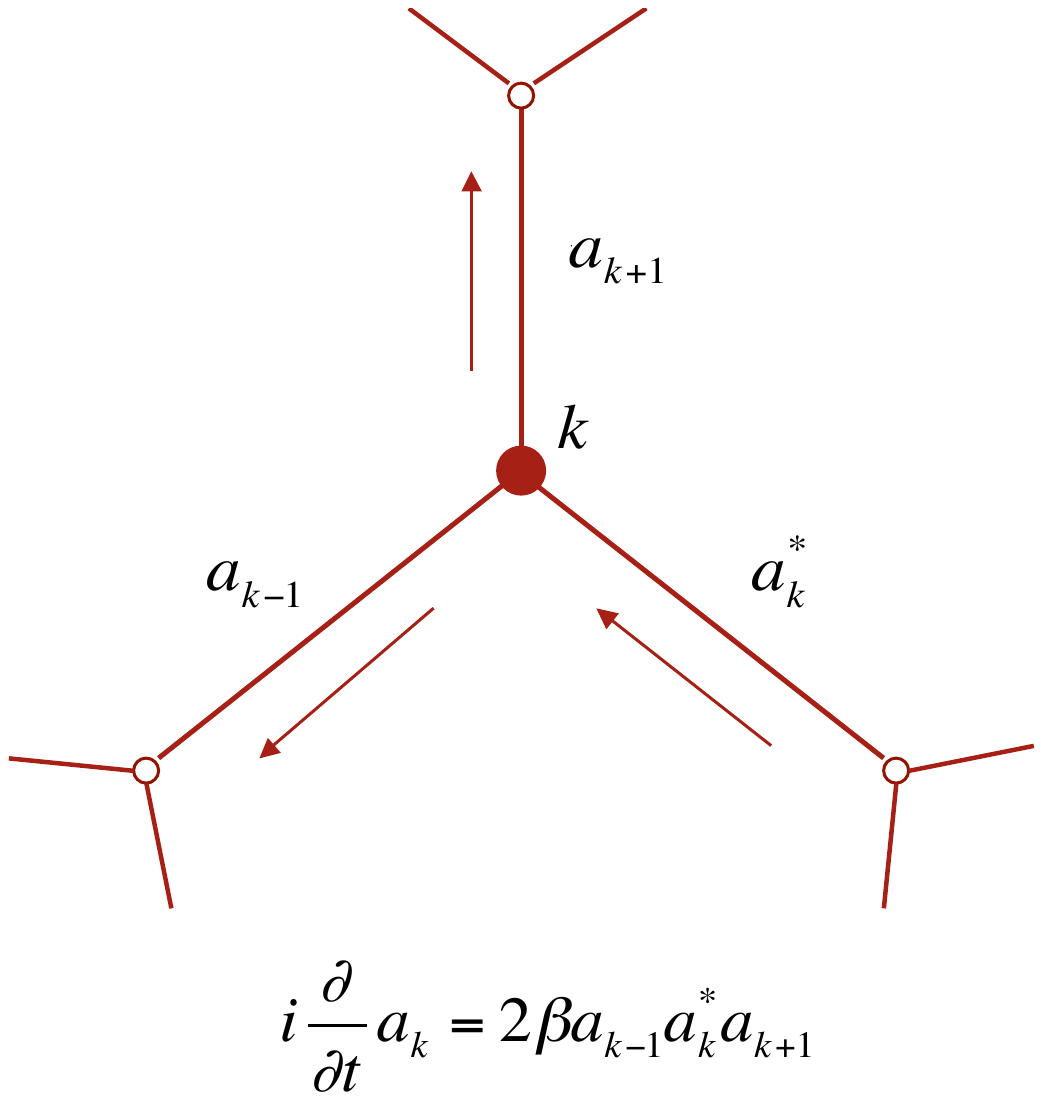}
\caption{\label{} Mapping dynamical equations~(\ref{9}) on a Cayley graph. 
}
\end{figure}

It is assumed, following Refs. \cite{EPL,PRE23}, that each nonlinear oscillator with the equation of motion~(\ref{9}) can reside in either chaotic (with probability $p$) or regular (with probability $1-p$) state. The probability for an oscillator to be in chaotic state is given by the Boltzmann factor $p = \exp (-\delta \omega / \Delta \omega_{\rm NL})$, where $\delta \omega \sim 1/\Delta n$ is the distance between Anderson eigenstates in the frequency domain, and $\Delta \omega_{\rm NL} = \beta |\psi_n|^2$ is the nonlinear frequency shift. Using here that $\sum_{n^\prime} |\psi_{n^\prime}|^{2} = 1$ implies $|\psi_n|^2 \sim 1/\Delta n$ for $\Delta n \gg 1$, one gets $p = \exp (-1/\beta)$ independently of $\Delta n$, that is $p$ is preserved through dynamics. 

If $p$ is not-too-close to zero yet smaller than the percolation threshold probability $p_c = 1/2$ \cite{Point}, then the chaotic states are clustered and do {\it not} constitute a percolation path throughout the field (and hence do not contribute to the d.c. conduction). If, however, hops can occur between the adjacent chaotic states only, and if this process occurs near its percolation threshold, then the dependence of a.c. conductivity over the frequency is given by the power law $\sigma_{\rm a.c.} (\omega) \propto \omega^\eta$, where $\eta = (\theta + d - d_f) / (2+\theta)$ \cite{Gefen,PRB01}, $d_f$ is the Hausdorff dimension of chaotic clusters, and $\theta$ is the connectivity index \cite{CommG}. Using known values $d=6$, $d_f = 4$, and $\theta = 4$ for mean-field percolation on Cayley trees \cite{Naka,Havlin}, one gets $\eta = 1$ exactly, hence $\sigma_{\rm a.c.} (\omega) \propto \omega$. This linear scaling extends the classic result of Elliott \cite{Elliott2,Elliott3} to DANSE~(\ref{DANSE}).    

Once the linear scaling $\sigma_{\rm a.c.} (\omega) \propto \omega$ is established, one repeats the renormalization procedure described above starting from the Kramers-Kronig transform~(\ref{KrKr}) to obtain, in place of DANSE~(\ref{DANSE}), the nonlinear Schr\"odinger equation with superquadratic nonlinearity, Eq.~(\ref{DSPSQ}), where the exponent of superquadratic power is written as $2s = 2(1+\mu_0)$, $\mu_0 = A_0/\pi\varepsilon_0 \ll 1$, and $A_0$ is the coefficient of a.c. conduction pertaining to the hopping motion. This type of nonlinearity steers automatically the nonlinear field to the margins of regularity via the dependence of the Chirikov overlap parameter $K\equiv \Delta\omega_{\rm NL} / \delta\omega \simeq \beta/|\Delta n|^{\mu_0}$ on the number of states. The end result is that the nonlinear field is asymptotically localized, with the nonlinear localization length    
\begin{equation}
\Lambda_{\rm loc} \sim \exp [(1/\mu_0) \ln \beta].
\label{Loc2} 
\end{equation} 
We define the range of validity of this result from the $p$ values lying beneath the percolation point\textemdash provided just that the chaotic states can form connected clusters\textemdash to the onset of percolation at $p_c = 1/2$, the percolation point included (because the d.c. conductivity is still zero for $p=p_c$ exactly \cite{Naka,Havlin}). The end result is that the nonlinear field of the type given by the DANSE~(\ref{DANSE}) is Anderson localized both at ($p= p_c$) and below ($p \lesssim p_c$) the percolation threshold. This is different from a conclusion drawn previously \cite{EPL} that the nonlinear field in Eq.~(\ref{DANSE}) is delocalized as soon as $p=p_c$. Here, we argue it is asymptotically localized through self-screening and the linear dependence $\sigma_{\rm a.c.} (\omega) \propto \omega$ in proximity to the percolation transition. Note that $\Lambda_{\rm loc} \sim \exp [(1/\mu_0) \ln \beta]$ can be greater than the lattice size, if $\mu_0 \ll 1$ is small enough. 

With use of $p = \exp (-1/\beta)$ one can translate $p_c = 1/2$ into a critical $\beta$ value yet permitting asymptotic localization: $\beta_c = 1/\ln 2$. For the $\beta$ values greater than this, an asymptotic localization is only possible through coupling to external medium. Within the localization length, one obtains the dispersion of field spreading along the lattice as the mean-squared displacement of the random walker on the incipient percolation cluster, leading to $\langle (\Delta n)^2 (t) \rangle \sim t^{2/(2+\theta)}$ \cite{Gefen,Havlin}. Using $\theta = 4$ for percolation on trees, one obtains $\langle (\Delta n)^2 (t) \rangle \sim t^{1/3}$. This scaling law has been observed numerically in Refs. \cite{Flach,Skokos,Many} and associated with weak chaos.

\acknowledgments
A.V.M. would like to thank the Isaac Newton Institute for Mathematical Sciences, Cambridge, U.K., for support and hospitality during the program ``Stochastic systems for anomalous diffusion," where work on this paper was undertaken. This work was supported by EPSRC grant No. EP/Z000580/1.

%
%
%
%


\begin{thebibliography}{}
\bibitem{Charge}
If the wave function $\psi_n$ refers to an electrically charged particle, then it is more convenient to calculate the effective $\beta$ value by looking into charges, not fields. One has ${\Delta\beta}/{\beta} \simeq {q_{\rm pol}}/{q_{0}}$, where ${q_{\rm pol}}$ is the induced polarization charge, and $q_{0}$ is the bare charge. Using for polarization charge ${q_{\rm pol}} = - q_0 \chi / (1+\chi)$, one gets ${\Delta\beta}/{\beta} \simeq - \chi/(1+\chi)$, leading to $\beta_{\rm eff} = \beta + \Delta \beta \simeq \beta / (1+\chi)$.
\bibitem{Comm}
In a basic theory of disordered dielectrics, one introduces \cite{Poll71,Elliott2} the following general form of a.c. conductivity specialized to hopping conduction, that is $\sigma_{\rm a.c.} (\omega) \propto \omega \int p(\tau) r^2 [{\omega\tau}/{(1+\omega^2\tau^2)}]d\tau$, where $p(\tau)$ is the probability distribution of relaxation times, $r^2$ derives from the square of the dipole moment, and the final term in square brackets is the Debye expression for dielectric loss. Because the random potential implies a spread of values (heights of the barriers), one has to integrate over $\tau$. For given $\omega$, the integral form above is dominated by the resonant contribution at $\omega\tau \sim 1$ (Tauberian theorem). Thus it follows that $\sigma_{\rm a.c.} (\omega)$ is a linear function of frequency, i.e., $\sigma_{\rm a.c.} (\omega) \propto \omega$ \cite{Elliott2,Elliott3}. In this context, any departures from linearity arise from the possibility that the hopping distance $r$ may be a function of $\omega$ \cite{Elliott2}.
\bibitem{DDI2}
The situation may be different in three dimensions (3D), where the inclusion of dipole-dipole interactions introduces spatial anisotropy into the dynamics (Ref. \cite{DDI}; references therein). In the limit of strong anisotropy, one may argue the existence of a preferred direction, along which one reduces to 1D the more general, 3D localization model. We ought to note that in three and higher dimensions, the Anderson localization is characterized by a mobility edge separating localized regime at low energy and diffusive states at high energy \cite{PW16}, and in this sense is fundamentally different from the 1D localization case, the subject matter of the present study.
\bibitem{Number}
The coordination number of a Cayley tree is the number of bonds at each node. The triad couplings in Eq.~(\ref{9}) dictate $z=3$.
\end{thebibliography}


\end{document}